\documentclass[pre,aps,amsmath,amssymb,reprint,floatfix,superscriptaddress,longbibliography]{revtex4-1}

\pdfoutput=1
\usepackage{xcolor}
\usepackage{siunitx}
\usepackage{graphicx}
\usepackage[colorlinks=true,citecolor=blue,urlcolor=blue]{hyperref}
\usepackage{etoolbox} 
\usepackage{bm} 
\usepackage[normalem]{ulem}
\newcommand{\new}[1]{\textcolor{black}{#1}}

\def\cal#1{\mathcal{#1}}
\def\eqq#1{Eq.~(\ref{#1})}

\def\eq#1{(\ref{#1})}

\def\f#1{Fig.~\ref{#1}}

\def\s#1{Section~\ref{#1}}

\def\c#1{~\cite{#1}}

\def\ccc#1{~Refs.~\cite{#1}}

\def\av#1{\langle #1 \rangle}

\def\beq{\begin{equation}}
\def\eeq{\end{equation}}
\def\bea{\begin{eqnarray}}
\def\eea{\end{eqnarray}}

\def\kB{k_{\rm B}}
\def\tf{t_{\rm f}}
\def\kt{\kB T}

\def\e{{\rm e}}

\begin{document}

\title{\new{Nonlinear} thermodynamic computing out of equilibrium}

\author{Stephen Whitelam}
\email{swhitelam@lbl.gov}

\author{Corneel Casert}
\email{ccasert@lbl.gov}
\affiliation{Molecular Foundry, Lawrence Berkeley National Laboratory, 1 Cyclotron Road, Berkeley, CA 94720, USA}

\begin{abstract}
We present the design for a thermodynamic computer that can perform arbitrary nonlinear calculations in or out of equilibrium. Simple thermodynamic circuits, fluctuating degrees of freedom in contact with a thermal bath and confined by a quartic potential, display an activity that is a nonlinear function of their input. Such circuits can therefore be regarded as thermodynamic neurons, and can serve as the building blocks of networked structures that act as thermodynamic neural networks, universal function approximators whose operation is powered by thermal fluctuations. We simulate a digital model of a thermodynamic neural network, and show that its parameters can be adjusted by genetic algorithm to perform nonlinear calculations at specified observation times, regardless of whether the system has attained thermal equilibrium. This work expands the field of thermodynamic computing beyond the regime of thermal equilibrium, enabling fully nonlinear computations, analogous to those performed by classical neural networks, at specified observation times.
\end{abstract}

\maketitle

{\em Introduction---} In classical forms of computing, thermal fluctuations are an obstacle to computation\c{bennett1982thermodynamics,landauer1991information,ceruzzi2003history}; for thermodynamic computing, thermal fluctuations are the means of doing computation\c{fry2005neural,conte2019thermodynamic,hylton2020thermodynamic,frank2017future,wolpert2019stochastic,fry2017physical,conte2019thermodynamic,wimsatt2021harnessing}. Fluctuations can drive state changes in devices, and can be used to encode information. For instance, consider a thermodynamic computer comprising scalar degrees of freedom $x_i$ that interact via the bilinear couplings $J_{ij} x_i x_j$. If this computer is placed in contact with a thermal bath at temperature $T$, then its equilibrium two-point correlations $\av{x_i x_j}_0 = \kt (J^{-1})_{ij}$ encode the elements of the matrix inverse of $J$. Thus measuring such correlations in equilibrium can be used to do matrix inversion\c{aifer2024thermodynamic,melanson2025thermodynamic}. 

The current focus of thermodynamic computing is to arrange for the equilibrium properties of a thermodynamic computer, described by the Boltzmann distribution, to correspond to the output of a specified computation. This approach is powerful because the potential energy of a thermodynamic computer specifies the Boltzmann distribution, and so by designing the potential we can design the computer's equilibrium properties\c{aifer2024thermodynamic}. However, this approach comes with two challenges. One is that we need the computer to attain thermal equilibrium. In general, physical systems equilibrate on a broad range of timescales\c{arceri2022glasses,biroli2013perspective,hagan2011mechanisms,whitelam2015statistical}, and the equilibration times for even a simple thermodynamic computer can vary by orders of magnitude as its program is altered\c{whitelam2025increasing}. A second challenge is that not every calculation can be represented by the Boltzmann distribution in an obvious way. For example, the matrix inversion described above can only be done if the matrix $J_{ij}$ is positive definite; if not, the system does not possess a well-defined equilibrium distribution.

\begin{figure*}[]
\centering
\includegraphics[width=\linewidth]{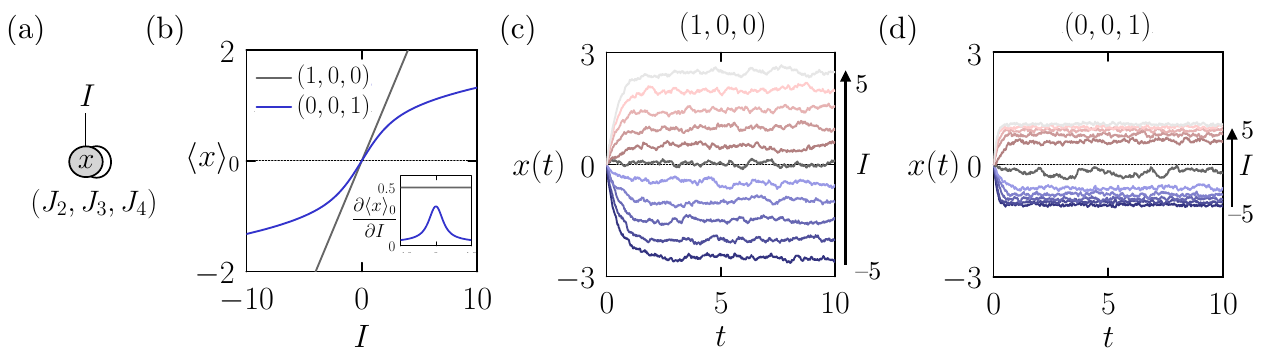}
\caption{(a) A thermodynamic circuit whose interaction energy is given by~\eqq{nrg} can function as a thermodynamic neuron. \new{The circle represents a scalar degree of freedom $x$. The curved line represents its intrinsic energy, the terms in ${\bm J}=(J_2,J_3,J_4)$ in \eqq{nrg}. The straight line represent an input signal or bias, the term in $I$ in \eqq{nrg}.} (b) Equilibrium activation function $\av{x}_0$ of the neuron, \eqq{act}, as a function of the neuron input $I$, for the case $\beta=1$. The vector ${\bm J}=(J_2,J_3,J_4)$ sets the values of the intrinsic couplings of the neuron. The quadratic-potential activation function is linear, while the quartic-potential activation function is nonlinear. (c) Dynamical evolution \eq{lang1} of the quadratic-potential neuron, for $\beta=100$, for 11 evenly-spaced values of $I$. (d) The same for the quartic-potential neuron. For times longer than some short threshold, the finite-time response is a nonlinear function of $I$.}
\label{fig1}
\end{figure*}

We can sidestep these challenges by arranging for a thermodynamic computer to perform calculations {\em out} of equilibrium. Out of equilibrium we lose contact with the theoretical foundation provided by the Boltzmann distribution, and so we must find other ways of programming a thermodynamic computer in order to do specified calculations. Some exceptions to the equilibrium paradigm already exist. For instance, the matrix exponential $e^{- J t}$ can be calculated at observation time $t$\c{duffield2023thermodynamic}, and nonequilibrium work measurements can be used to calculate the determinant of a matrix\c{aifer2024thermodynamic}. However, no design exists for \new{a thermodynamic computer that operates at specified observation times and is programmable in a general sense, meaning that it is capable of approximating arbitrary continuous functions. Here we provide such a design by introducing a thermodynamic computer that is analogous to a neural network.} A thermodynamic computer of this nature is a nonlinear model that can serve as a universal function approximator, and can be programmed to perform arbitrary nonlinear computations at specified observation times. This is true whether or not the computer has attained thermodynamic equilibrium at those observation times. 

\new{The equilibrium thermodynamic computers of \ccc{aifer2024thermodynamic, melanson2025thermodynamic}, which motivate this work, can be viewed as continuous-spin analogs of Hopfield networks\c{hopfield1982neural} or Boltzmann machines\c{hinton2017boltzmann,salakhutdinov2009deep}, statistical mechanical models that represent probability distributions over binary variables. Since a Boltzmann machine encodes information in its equilibrium or Boltzmann distribution, one could refer to the models of \ccc{aifer2024thermodynamic, melanson2025thermodynamic} as {\em Boltzmann computers}: they are designed to sample from the Boltzmann equilibrium corresponding to the system's potential energy function. In such models, the specific form of the dynamics, whether overdamped or underdamped, is less important than the fact that it is microscopically reversible and converges to the desired equilibrium.}

\new{Our contribution is to explore the computational capabilities of such systems when extended to include nonlinear neuron potentials and operated out of equilibrium. Specifically, we encode the outcome of a computation in the dynamical trajectories of the system, without appealing to the Boltzmann distribution. A computer operating in this mode, using Langevin trajectories to perform a computation, could be termed a {\em Langevin computer}.}

\new{The theoretical framework describing thermodynamic computing includes nonequilibrium statistical mechanics, stochastic thermodynamics, and information theory. These fields offer a set of tools for analyzing small, fluctuating systems, such as fluctuation theorems, thermodynamic speed limits, and uncertainty relations\c{seifert2012stochastic,wolpert2024stochastic,sagawa2014thermodynamic}. For example, the Jarzynski equality, combined with nonequilibrium work measurements, can be used to compute the determinant of a matrix\c{aifer2024thermodynamic}. Thermodynamic computers often use added noise to accelerate computational, and the energetic cost of this addition can be quantified within stochastic thermodynamics\c{whitelam2025increasing}. Learning rules can be linked to thermodynamic quantities: for instance, a generative thermodynamic computer trained by gradient descent minimizes heat dissipation during training\c{whitelam2025generative}.}

{\em Thermodynamic neurons ---} In more detail, we introduce the thermodynamic circuit shown in \f{fig1}(a), a fluctuating classical degree of freedom placed in contact with a heat bath and confined by a quartic potential~\footnote{In general, the confining potential need not be exactly quartic, but it must be higher-order than quadratic and thermodynamically stable \new{(i.e. bounded from below)}. We can consider \eqq{nrg} to represent a Maclaurin expansion, in powers of $x$, of an arbitrary nonlinear thermodynamic circuit.}. This circuit represents a scalar degree of freedom $x$ that experiences the potential energy
\beq
\label{nrg}
U_{\bm J}(x,I) = J_2 x^2+J_3 x^3+J_4 x^4 -Ix.
\eeq
The parameters ${\bm J}=(J_2,J_3,J_4)$ are the intrinsic couplings of the circuit, and $I$ is an input signal. We can consider the circuit to represent a thermodynamic neuron, whose activation function is the relation between the output $x$ and the input $I$. The output must be a nonlinear function of the input in order for a network built from such neurons to be a universal approximator.

Let the neuron be put in contact with a thermal bath at temperature $T$. In thermal equilibrium, the output of the neuron has the mean value
\beq
\label{act}
\av{x}_0 \equiv \frac{ \int {\rm d} x\, x \, \e^{-\beta U_{\bm J}(x,I)}}{ \int {\rm d} x \,\e^{-\beta U_{\bm J}(x,I)}}.
\eeq
When the neuron potential is purely quadratic, i.e. ${\bm J}=(J_2,0,0)$, the integrals in \eq{act} can be solved analytically, giving the linear form $m=I/(2 J_2)$. This form is plotted, for $J_2=1$, as a gray line in \f{fig1}(b); the horizontal dotted black line denotes the value zero. In this case the equilibrium activation function of the neuron is linear, meaning that networks of such neurons in equilibrium cannot serve as universal function approximators\c{cybenko1989approximation,hornik1989multilayer}. The simplest case that is thermodynamically stable \new{(i.e. the potential is bounded from below)} and admits a nonlinear equilibrium activation function is the purely quartic case, ${\bm J} = (0,0,J_4)$, with $J_4>0$. This case is shown as a blue line in \f{fig1}(b), with $J_4=1$~\footnote{In the purely quartic case the equilibrium activation function can be expressed analytically, in terms of the hypergeometric function.}. The equilibrium activation function is nonlinear: its gradient (shown inset) is largest near the origin, and decreases as $|I|$ becomes large. 

An additional design consideration is the variance $\sigma^2=\av{x^2}_0-\av{x}_0^2$ of the neuron's equilibrium fluctuations. We show in the Supplementary Material that adding a quadratic term to the quartic term suppresses the neuron's fluctuations in equilibrium.  The larger the fluctuations of a thermodynamic neuron's output, the more samples will be required to compute a meaningful signal when observing a computer built from such neurons. For these reasons we choose our default neuron parameters to be ${\bm J}=(1,0,1)$: the quartic coupling induces nonlinearity, while the quadratic coupling serves to suppress fluctuations near $I=0$. 

The nonequilibrium response of the quartic-potential thermodynamic neuron is also, in general, nonlinear with input. Thermodynamic computers operate under Langevin dynamics\c{aifer2024thermodynamic,melanson2025thermodynamic}, and here we consider the overdamped dynamics
\beq
\label{lang1}
\dot{x}_i= -\mu \frac{\partial V({\bm x},{\bm I})}{\partial x_i}  + \sqrt{2 \mu \kt} \, \eta_i(t),
\eeq 
where $i$ labels the neuron; $V({\bm x},{\bm I})$ is the computer's potential given input ${\bm I}$; and the Gaussian white noise terms satisfy $\av{\eta_i(t)}=0$ and $\av{\eta_i(t) \eta_j(t')} = \delta_{ij} \delta(t-t')$. The mobility parameter $\mu$ sets the basic time constant of the computer. For the thermodynamic computers of Refs.\c{aifer2024thermodynamic,melanson2025thermodynamic}, $\mu^{-1} \sim 1$ microsecond. For damped oscillators made from mechanical elements\c{dago2021information} or Josephson junctions\c{ray2023gigahertz}, $\mu^{-1}$ is of order a millisecond or a nanosecond, respectively. 

In \f{fig1}(c,d) we show the dynamical evolution of a single neuron $x$ in the potential $V({\bm x},{\bm I})=U_{\bm J}(x,I)$, under the dynamics \eq{lang1}, starting from $x=0$. Panel (c) shows the case of the quadratic potential ${\bm J}=(1,0,0)$, for various fixed values of the input $I$. The neuron is initially out of equilibrium, converging to equilibrium in about 2 time units (time is expressed in units of $\mu^{-1}$). As described by \f{fig1}(b), the mean value of $x$ in equilibrium is a linear function of $I$. 

Panel (d) shows the case of a quartic potential ${\bm J}=(0,0,1)$. For observation times larger than some threshold, the neuron's output is a nonlinear function $x(I)$, and eventually converges to the equilibrium activation function $\av{x(I)}_0$, which we have designed to be nonlinear. A network of such neurons, observed on similar timescales, can therefore function as a universal approximator. \new{(In a related vein, nonlinear physical neural networks that operate on the energy scales of classical computing have recently been trained to do nonlinear computations\c{dillavou2024machine}.)}

  \begin{figure*}[]
\centering
\includegraphics[width=\linewidth]{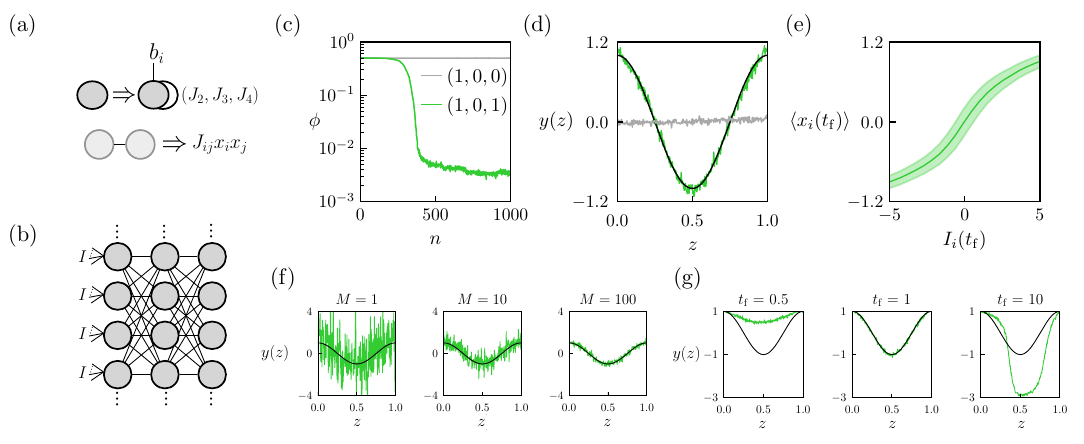}
\caption {(a,b) Elements of a thermodynamic computer analogous to a neural network. (a) The thermodynamic neurons described in~\f{fig1} are connected by bilinear couplings. \new{Top: a single circle implies a neuron $x_i$ of the type described in \f{fig1}, with an input (bias) $I=b_i$ and a nonlinear potential parameterized by ${\bm J}=(J_1,J_2,J_3)$. Bottom: lines between circles imply a bilinear coupling $J_{ij} x_i x_j$.} (b) \new{With the visual shorthand described in panel (a),} we consider layered networks of such neurons, with adjacent layers coupled all-to-all, having total potential energy~\eq{pot_tot}. (c) Training a simulation model of a thermodynamic computer to express a nonlinear function at a specified observation time. We show loss \eq{phi} as a function of evolutionary time $n$ for a layered thermodynamic computer built from quadratic neurons (gray) or quadratic-quartic neurons (green). (d) Output \eq{out} at observation time $\tf=1$ of the linear computer (gray) and the nonlinear computer (green), as a function of the input $z$, averaged over $M=10^3$ samples. The target function is shown as a black line. (e) Mean neuron activations measured at observation time $\tf$ as a function of the neuron inputs at the same time, for the nonlinear model. The color band denotes $\pm$ one standard deviation. (f) Output~\eq{out} at time $\tf=1$ of the trained nonlinear thermodynamic computer as a function of input $z$, computed using $M$ samples. The target function is shown as a black line (training was done using $M=10^3$ samples). (g) Output~\eq{out} at various observation times $\tf$ of the trained nonlinear thermodynamic computer, as a function of input $z$, computed using $M=10^3$ samples. The target function is shown as a black line. The computer is trained so that it reproduces the target function when observed at time $\tf=1$.}
\label{fig2}
\end{figure*}

{\em Thermodynamic neural network ---} To illustrate this statement, consider a thermodynamic computer consisting of a network of $N$ thermodynamic neurons $x_i$, with potential energy function
\beq
\label{pot_tot}
V({\bm x},{\bm I})=V_{\rm int}({\bm x})+V_{\rm ext}({\bm x},{\bm I})
\eeq
Here
\beq
\label{pot}
V_{\rm int}({\bm x}) = \sum_{i=1}^N U_{\bm J}(x_i,b_i)+\sum_{(ij)} J_{ij} x_i x_j.
\eeq
is the internal potential of the computer, accounting for neuron self-interactions and connections between neurons. The first sum in \eq{pot} runs over $N$ single-neuron energy terms \eq{nrg}, while the second sum runs over all distinct pairs of connected neurons. We use the bilinear interaction of Refs.\c{aifer2024thermodynamic,melanson2025thermodynamic}. The computers described in those papers use an all-to-all coupling; here, to make contact with existing neural-network designs, we consider the layered structure shown in \f{fig2}, with all-to-all connections between layers. This design mimics that of a conventional deep fully-connected neural network. However, unlike in a conventional deep neural network, in which information flows from the input layer to the output layer, the bilinear interaction $J_{ij} x_i x_j$ ensures that neuron $i$ communicates with neuron $j$, and vice versa, and so information flows forward {\em and} backward between the layers of the thermodynamic computer.

To provide input to the thermodynamic computer we introduce the external coupling
\beq
\label{pot2}
V_{\rm ext}({\bm x},{\bm I}) =\sum_{{\rm inputs} (ij)} W_{ij} I_i x_j,
\eeq
where the sum runs over all connections between the external inputs $I_i$ and the input neurons $x_j$ (here the top-layer neurons), mediated by the parameters $W_{ij}$. 

The computer evolves according to the Langevin dynamics \eq{lang1}. Because the computer is noisy, we wish to take $M$ samples of its output and average over these samples. In the main text we do this by reset-sampling: starting from zero neuron activation ${\bm x} = {\bm 0}$, we run the computer for time $\tf$, observe the outcome, reset the neuron activations to zero, and repeat the procedure, gathering $M$ samples in total (in the Supplementary Material we also consider the case of serial sampling). Reset sampling naturally lends itself to parallelization: the $M$ samples can be computed independently, on distinct copies of the thermodynamic computer if such copies are available. Reset sampling can also be done using a single computer whose neurons are reset periodically.

We designate the final-layer neurons of the computer as its outputs, and calculate the reset-sampling averages
\beq
\label{s1}
\av{x_i({\bm I})}_{\rm r}= M^{-1}\sum_{\alpha=1}^M x^{(\alpha)}_i({\bm I},\tf),
\eeq
where \new{the subscript `r' indicates `reset-sampling', and} the sum runs over $M$ independent \new{trajectories} $\alpha$ of the dynamics \eq{lang1}. The only requirement on $\tf$ is that it is long enough that the effective activation function of the neuron is nonlinear. We set $\tf=1$, in units of $\mu^{-1}$.

{\em Programming a thermodynamic computer---} A nonlinear thermodynamic computer can be programmed to perform arbitrary nonlinear computations at specified times. We first consider the task of expressing a nonlinear function of a single variable. In this case the computer has one input, $I=z$. We define the target function $y_0(z) \equiv \cos( 2 \pi z)$ and the loss function 
\beq
\label{phi}
\phi  \equiv K^{-1} \sum_{j=1}^K \left(y_0(z_j)-y(z_j)\right)^2,
\eeq 
where the sum runs over $K=250$ evenly-spaced points $z_j=j/(K-1)$ on the interval $z \in [0,1]$. The quantity
\beq
\label{out}
y(z)\equiv \sum_{i \in {\rm outputs}} f_i \av{x_i(z)}_{{\rm r},{\rm s}}
\eeq
is the output of the thermodynamic computer, given the input $z$, averaged over $M=10^3$ samples. \new{The subscript `r,s' indicates that averages are taken either in reset-sampling or in serial-sampling mode. If the former, each of the $M$ samples is obtained from an independent trajectory. If the latter, all $M$ samples are obtained from a single trajectory.}

The adjustable parameters of the computer are ${\bm \theta}= \{W_{ij}\} \cup \{b_i\} \cup \{J_{ij}\} \cup \{f_i\}$. Here $\{W_{ij}\}$ is the set of input weights specified by \eqq{pot2}; $\{b_i\}$ is the set of biases specified by~\eqq{pot}; $\{J_{ij}\}$ is the set of connections specified by the same equation; and $\{f_i\}$ is a set of weights that couple to the output neurons. To program the computer we adjust the parameters ${\bm \theta}$ using a genetic algorithm instructed to minimize a loss function $\phi$, using an efficient GPU implementation (see Supplementary Material). 

Following training, the identity of the computer's parameters are fixed, and the computer can be run for any chosen input. The parameters of the digital model of the thermodynamic computer could in principle be implemented in hardware, resulting in a device designed to output a specified computation at a specified time, powered by thermal fluctuations. We note that if the hardware implementation is not an exact copy of the digital model, the genetic-algorithm training could be continued directly in hardware: the procedure can be applied to an experimental system exactly as it is applied to a simulation model\c{sabattini2024adaptive}. 

We choose a layered computer design of width 8 and depth 4. We consider two types of thermodynamic neuron: a quadratic-quartic thermodynamic neuron, ${\bm J}=(1,0,1)$, which gives rise to a nonlinear computer, and a quadratic thermodynamic neuron, ${\bm J}=(1,0,0)$, which gives rise to a linear computer. We take $\beta=10$, so that the neuron energy scale is 10 times that of the thermal energy.

In \f{fig2}(c) we show the loss as a function of evolutionary time for the two models. The linear model fails to train -- it cannot express a nonlinear function of the input variable -- while the nonlinear model learns steadily, reaching a small value of the loss. Panel (d) shows the output functions learned by the two models: the nonlinear model has learned a good approximation of the target cosine function. The intrinsic noise of the computer is visible in the output, but for $M=10^3$ samples, for each value of $z$, the mean output signal of the computer exceeds the scale of the noise by a considerable margin.

Panel (e) of \f{fig2} shows the sampled neuron outputs as a function of the neuron inputs (the inputs being all signals in to the neuron, excepting the thermal noise) at the designated observation time. The nonlinear model possesses a nonlinear finite-time activation function, explaining the computer's ability to learn an arbitrary nonlinear function. By contrast, the quadratic-neuron computer is at all times a linear model, and is unable to express a nonlinear function.

Training is done using $M=10^3$ samples for each value of the input $z$, but the trained computer can be used with fewer samples if desired. In \f{fig2}(f) we show the output of the trained thermodynamic computer, as a function of the input $z$, for a range of values of $M$.

Training in reset-sampling mode results in a thermodynamic computer programmed to express the target function at a prescribed observation time $\tf=1$. In \f{fig2}(g) we show the output of the computer at a range of observation times. The output of the computer varies as a function of time, and is equal to the target function only at the prescribed observation time. The output of the computer in equilibrium (corresponding to the long-time limit) is considerably different to the target function. In this example, therefore, the programmed thermodynamic computer operates far from equilibrium.

{\em Machine learning with a thermodynamic computer ---} Having confirmed the ability of a network of nonlinear thermodynamic neurons to express an arbitrary nonlinear function, we now consider a standard benchmark in machine learning, classifying the MNIST data set\c{lecun1998gradient}. MNIST consists of greyscale images of $70,000$ handwritten digits on a grid of $28 \times 28$ pixels, each digit belonging to one of ten classes $C \in [0,9]$. 

\begin{figure}[]
\centering
\includegraphics[width=\linewidth]{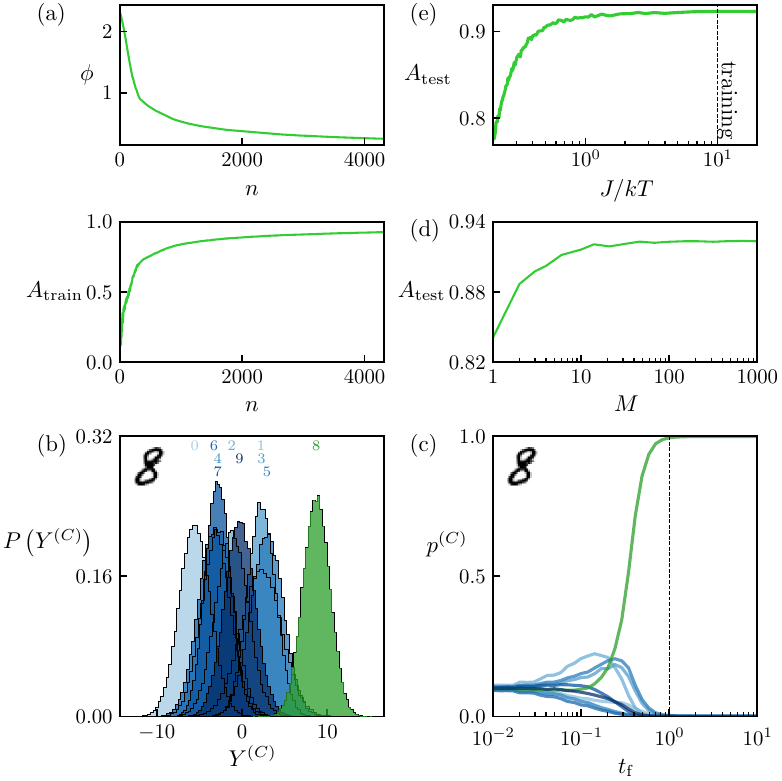}
\caption {Training a simulation model of a thermodynamic computer to classify MNIST. The computer, which consists of a 3-layer network of quadratic-quartic $(1,0,1)$ neurons, is trained in reset-sampling mode, using $M=10^3$ samples taken at observation time $\tf=1$. (a) Loss (cross-entropy) and training-set classification accuracy as a function of evolutionary time $n$. (b) For a single digit, an 8, we show the probability distribution, taken over $10^5$ samples, of the computer's per-sample class score. The mean value of each distribution, which is the value used for classification, is indicated at the top of the panel. The correct distribution is shown in green, the others in shades of blue. (c) The class probabilities of the computer, upon being shown the indicated digit, for various observation time $\tf$. The computer is trained to classify the digit at an observation time $\tf=1$ (vertical dotted line). (d) Test-set classification accuracy of the trained computer, as a function of the number of samples $M$ generated by the computer (each taken at observation time $\tf=1$). (e) Test-set accuracy of the trained computer at a range of energy scales. The computer is trained at the energy scale $J=10 \kt$.}
\label{fig3}
\end{figure}

We simulate a 3-layer thermodynamic computer with quadratic-quartic $(1,0,1)$ neurons. Each layer has 32 neurons. Each neuron in the input layer couples to all the pixels $I_i$ of an MNIST digit via~\eqq{pot2}. The output layer of 32 neurons is used to construct the computer's prediction for the class of MNIST digit ${\bm I}_j$, and we train the computer by genetic algorithm, instructing it to minimize the cross-entropy between the class probabilities predicted by the thermodynamic computer and the ground-truth labels (see Supplementary Material).

In \f{fig3}(a) we show the loss as a function of evolutionary time $n$ as the computer is trained to classify MNIST. The computer learns steadily under the action of the genetic algorithm. Panel (b) shows the corresponding training-set classification accuracy (which can be observed but is not used during training). The corresponding test-set accuracy after training is about 93\%, which is not state-of-the-art -- many other methods classify MNIST with greater accuracy\c{mnist_leaderboard} -- but it is more accurate than a linear classifier, and confirms the ability of a thermodynamic computer to address standard machine-learning problems. As with conventional neural networks, better accuracy will be achieved using different computer designs and methods of training. Here our aim is to show proof of principle: if implemented in hardware, this thermodynamic computer would be able, powered only by thermal fluctuations, to classify MNIST digits.

In \f{fig3}(b) we show the output of the trained computer when presented with a single digit, an 8, which it correctly classifies. \new{We distinguish between the class score $Y^{(C)}$, the raw output of the thermodynamic computer prior to normalization, and the class probability $p^{(C)}$, which results from applying a softmax transformation to averaged score. We use the term class prediction to refer to the class with the highest score, i.e., the computer's predicted label for the digit. We plot the probability distribution, taken over $10^5$ samples, of the computer's class score $Y^{(C)}$ for the digit. The mean value of each distribution (the value used for classification) is indicated at the top of the panel. } The correct distribution is shown in green, with the others in shades of blue. In panel (c) we plot the value of the \new{trained computer's 10 class probabilities} upon being shown the indicated digit. The computer is trained to classify the digit at an observation time $\tf=1$. In this case the computer has attained a steady-state dynamics at the specified observation time, but this is not a general phenomenon: when presented with other digits, the computer's neurons are still evolving at $\tf=1$, and so in general the computer operates out of equilibrium.

\f{fig3}(d) shows the test-set accuracy of the computer upon collecting $M$ samples. Training was done with $M=10^3$ samples, but similar accuracy can be obtained using only about 20 samples. Thus if implemented in hardware, the thermodynamic computer could perform classification relatively efficiently.

The ability of a trained computer to operate at different noise scales depends on the nature of the problem it is trained for and its architecture. In \f{fig3}(e) we show the test-set accuracy of the trained thermodynamic computer at a range of energy scales, calculated by holding fixed the parameters of the thermodynamic computer and varying the noise strength (see Supplementary Material). The computer, trained at the energy scale $J=10 \kt$, performs essentially as well when subjected to noise comparable to its own energy scale ($J=\kt$). This result shows the ability of the computer to operate reliably throughout the regime characteristic of thermodynamic computing $(\kt \lesssim J \ll 10^3 \kt)$, and shows its output to be robust to small changes in noise level (which might occur if a device becomes hot during computation). For sufficiently large noise levels the computer's performance begins to decline.

{\em Conclusions---} Classical computing aims to suppress thermal fluctuations, while thermodynamic computing uses them. Most existing thermodynamic computing focus on doing linear algebra in equilibrium, using the Boltzmann distribution to encode computations\c{aifer2024thermodynamic,melanson2025thermodynamic}. \new{A thermodynamic computer used in this way could be described as a {\em Boltzmann computer}}. However, equilibration can be slow, and not all problems \new{can be encoded as a Boltzmann distribution}. This paper presents an alternative approach, in which a nonlinear thermodynamic computer is trained to perform nonlinear calculations at specified times. \new{A thermodynamic computer trained to perform computations using Langevin trajectories could be called a {\em Langevin computer.}} 

 The core component of the design, a thermodynamic neuron, is a fluctuating degree of freedom confined by a non-quadratic potential, allowing networks of such neurons to act as universal approximators. \new{A recent paper\c{lipka2024thermodynamic} presented the design for a thermodynamic neuron realized by qubits coupled to multiple thermal baths. The design presented here is based on a fully classical model of a nonlinear thermodynamic neuron, and requires only a single thermal bath.} Our proposed design, \new{sketched in \f{fig2}, could be implemented using existing hardware elements: the neurons could be realized by RLC\c{melanson2025thermodynamic} circuits with nonlinear components, or by superconducting circuits with Josephson junctions\c{ray2023gigahertz}}.  A thermodynamic computer of this nature could be programmed to perform arbitrary nonlinear computations at specified times, powered by thermal fluctuations.
 
{\em Acknowledgments---} This work was done at the Molecular Foundry, supported by the Office of Science, Office of Basic Energy Sciences, of the U.S. Department of Energy under Contract No. DE-AC02-05CH11231. This research used resources of the National Energy Research Scientific Computing Center (NERSC), a U.S. Department of Energy Office of Science User Facility located at Lawrence Berkeley National Laboratory, operated under Contract No. DE-AC02-05CH11231. C.C. was supported by a Francqui Fellowship of the Belgian American Educational Foundation, and by the US DOE Office of Science Scientific User Facilities AI/ML project ``A digital twin for spatiotemporally resolved experiments''.


%

\section*{Supplementary Material}

\renewcommand{\theequation}{S\arabic{equation}}
\renewcommand{\thefigure}{S\arabic{figure}}
\renewcommand{\thesection}{S\arabic{section}}

\setcounter{equation}{0}
\setcounter{section}{0}
\setcounter{figure}{0}

\section{Overview}

In this supplement we describe the design and training of the nonlinear thermodynamic computer presented in the main text. The core of the design is a thermodynamic neuron, a fluctuating classical degree of freedom placed in contact with a heat bath and confined by a quartic potential~\footnote{In general, the confining potential need not be exactly quartic, but it must be higher-order than quadratic and thermodynamically stable.}. The equilibrium average of the neuron activation is a nonlinear function of the signal input to it, meaning that a network built from interacting thermodynamic neurons can function as a universal approximator\c{cybenko1989approximation,hornik1989multilayer}. Within a digital simulation of a thermodynamic computer we construct networks of such neurons using the bilinear interactions characteristic of existing thermodynamic computers\c{aifer2024thermodynamic,melanson2025thermodynamic}. We train these computers by genetic algorithm to perform nonlinear computations -- expressing a nonlinear function and classifying MNIST -- at specified times. Thermodynamic computers of this nature could be considered to be thermodynamic neural networks, or ``thermoneural networks''.

This results of this paper expand the field of thermodynamic computing beyond linear algebra and thermodynamic equilibrium, enabling fully nonlinear computations, comparable to those performed by conventional neural networks, at specified observation times. The design and training of the computer are done by `digital twin', applying a genetic algorithm to a simulation model of the thermodynamic computer realized on a classical digital computer. The elements of the present design have been implemented in hardware\c{aifer2024thermodynamic,melanson2025thermodynamic}, with the exception of the quartic neuron potential. If the latter can be engineered -- perhaps by using nonlinear inductors or capacitors to induce quartic self-interactions within RLC circuits, or using the nonlinear inductance provided by Josephson junctions\c{ray2023gigahertz} -- then the resulting nonlinear computer can be programmed, e.g. by genetic algorithm. The result would be a thermodynamic computer that would be driven by thermal fluctuations to perform a specified computation at a specified observation time, whether or not the computer has come to equilibrium.

This work adds to the literature of classical models that can be used to perform calculations, such as Hopfield networks\c{hopfield1982neural}, Boltzmann Machines\c{ackley1985learning}, physical neural networks\c{stern2021supervised}, and charge-based thermodynamic neural networks that self-organize under external drive\c{hylton2020thermodynamic}. Our design is analogous to conventional perceptron-based neural networks in that it is programmable and can function as a universal approximator, but differs in that it is stochastic and designed to be implemented in hardware, where its operation would be driven by the natural dynamics of fluctuating classical degrees of freedom. A recent paper\c{lipka2024thermodynamic} presented the design for a thermodynamic neuron realized by qubits coupled to multiple thermal baths. The design presented here is based on a fully classical model of a nonlinear thermodynamic neuron, and requires only a single thermal bath. 

In \s{supp_neuron} we introduce and analyze the properties of a simple thermodynamic circuit that can function as a thermodynamic neuron, a fluctuating degree of freedom whose output is a nonlinear function of its input. In \s{supp_net} we simulate interacting networks of thermodynamic neurons, and train them by genetic algorithm to perform specified nonlinear computations.

\section{Thermodynamic neurons}
\label{supp_neuron}

\begin{figure*}[]
\centering
\includegraphics[width=\linewidth]{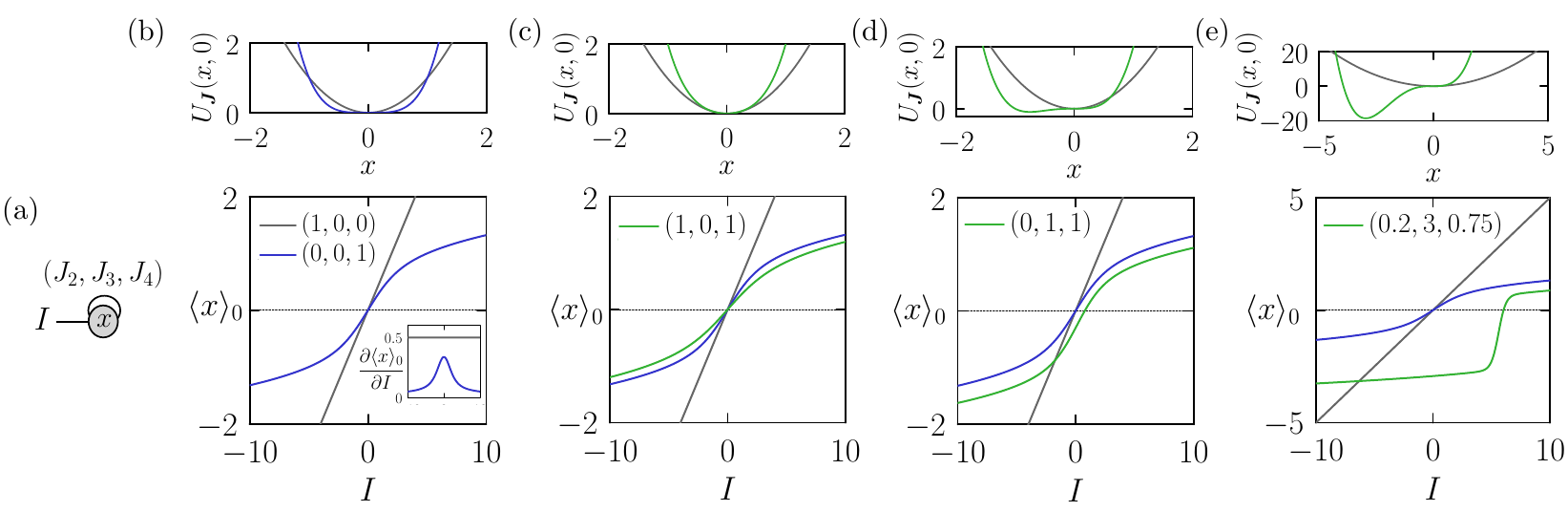}
\caption{(a) A thermodynamic circuit whose interaction energy is given by~\eqq{supp_nrg} can function as a thermodynamic neuron. In panels (b--e) we show the equilibrium activation function $\av{x}_0$ of the neuron, \eqq{supp_act}, as a function of the neuron input $I$, for the case $\beta=1$. The vector ${\bm J}=(J_2,J_3,J_4)$ sets the values of the intrinsic couplings of the neuron. The top panels in (b--e) show the potential \eq{supp_nrg} at zero input, for the quadratic case (gray) and the case introduced in the lower panel (blue or green). (b) Purely quadratic (gray) and quartic (blue) cases. The purely quartic case (with $J_4>0$) is the simplest case that is thermodynamically stable and admits a nonlinear activation function. Both cases are shown for reference in the following panels. Inset: gradient of activation function on the same horizontal scale as the main panel. (c) Example in which a quadratic coupling is included with the quartic coupling. (d) Example in which a cubic coupling is included with the quartic coupling. (e) Example in which all three couplings are nonzero. }
\label{supp_fig1}
\end{figure*}

\subsection{Equilibrium behavior} 

To motivate the construction of a thermodynamic computer analogous to a neural network, consider the thermodynamic circuit shown in \f{supp_fig1}(a). This circuit represents a scalar degree of freedom $x$ that experiences the potential energy
\beq
\label{supp_nrg}
U_{\bm J}(x,I) = J_2 x^2+J_3 x^3+J_4 x^4 -Ix.
\eeq
The parameters ${\bm J}=(J_2,J_3,J_4)$ are the intrinsic couplings of the circuit, and $I$ is an input signal. We can consider the circuit to represent a thermodynamic neuron, whose activation function is the relation between the output $x$ and the input $I$. The output must be a nonlinear function of the input in order for a network built from such neurons to be a universal approximator.

Let the neuron be put in contact with a thermal bath at temperature $T$. Our ultimate goal is to consider networks of such neurons operating at specified observation times, but an important design consideration in pursuit of this goal is the equilibrium behavior of a single neuron. In thermal equilibrium, the output of the neuron has the mean value
\beq
\label{supp_act}
m=\av{x}_0,
\eeq
where
\beq
\label{supp_act2}
\av{\cdot}_0 \equiv \frac{ \int {\rm d} x\, (\cdot) \, \e^{-\beta U_{\bm J}(x,I)}}{ \int {\rm d} x \,\e^{-\beta U_{\bm J}(x,I)}}.
\eeq
When the neuron potential is purely quadratic, i.e. ${\bm J}=(J_2,0,0)$ -- with $J_2>0$ to ensure thermodynamic stability -- the integrals in \eq{supp_act2} can be solved analytically, giving the linear form $m=I/(2 J_2)$. This form is plotted, for $J_2=1$, as a gray line in \f{supp_fig1}(b); the horizontal dotted black line denotes the value zero. In this case the equilibrium activation function of the neuron is linear, meaning that networks of such neurons in equilibrium cannot serve as universal function approximators~\footnote{The quadratic nonlinearity also gives rise to a linear activation function out of equilibrium.}.
  \begin{figure}[]
\centering
\includegraphics[width=\linewidth]{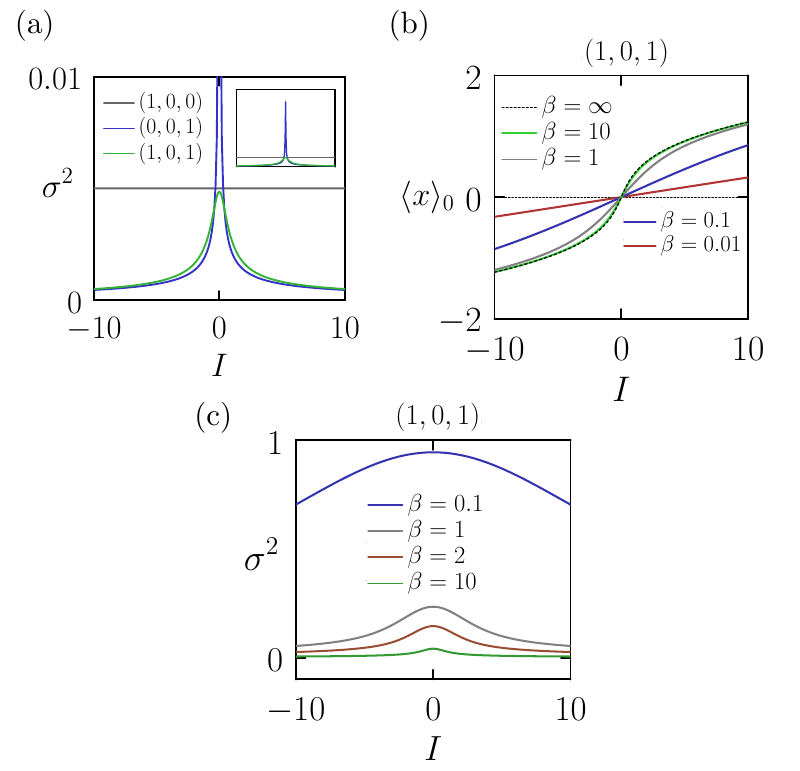}
\caption{(a) Equilibrium fluctuations \eq{supp_var} of the thermodynamic neuron of \f{supp_fig1}(a), for the case $\beta=100$. The addition of the quadratic coupling to the quartic one (green), suppresses fluctuations relative to the pure quartic case (blue). The {\em mean} equilibrium activation functions of those two cases are similar; see \f{supp_fig1}(c). The inset shows the largest fluctuations of the purely quartic neuron to be many times that of the purely quadratic neuron. (b,c) The mean (b) and variance (c) of the equilibrium activation function of the $(1,0,1)$ neuron depend on temperature.}
\label{supp_fig2}
\end{figure}

 \begin{figure*}[]
\centering
\includegraphics[width=\linewidth]{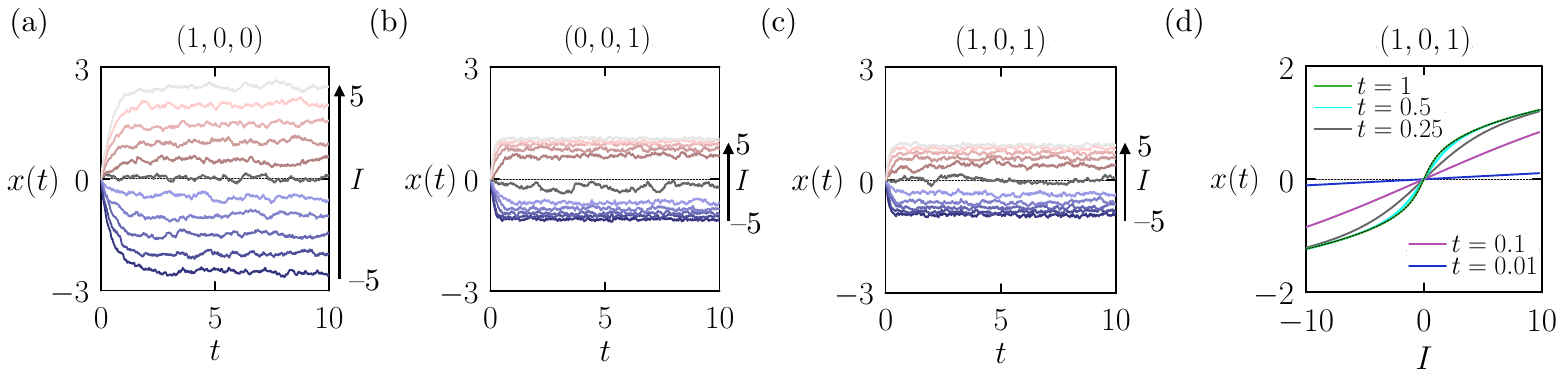}
\caption{Nonequilibrium properties of the thermodynamic neuron of \f{supp_fig1}(a). (a--c) Langevin evolution \eq{supp_lang1} of the neuron $x$ for $\beta=100$, for 11 evenly-spaced values of $I$. We consider three different sets of couplings ${\bm J}$: (a) purely quadratic; (b) purely quartic; and (c) mixed quadratic-quartic. (d) The finite-time activation function of the (1,0,1) neuron under the dynamics \eq{supp_lang1} (here for $\beta=\infty$) is nonlinear above some threshold observation time (here about 0.2 time units), and for longer times converges to the equilibrium result (curved black dashed line).}
\label{supp_fig3}
\end{figure*}

The simplest case that is thermodynamically stable and admits a nonlinear equilibrium activation function is the purely quartic case, ${\bm J} = (0,0,J_4)$, with $J_4>0$. This case is shown as a blue line in \f{supp_fig1}(b), with $J_4=1$~\footnote{In the purely quartic case the equilibrium activation function can be expressed analytically, in terms of the hypergeometric function.}. The equilibrium activation function is nonlinear: its gradient (shown inset) is largest near the origin, and decreases as $|I|$ becomes large. The gradient remains finite for finite $I$: the neuron does not saturate. 

Panels (c), (d), and (e) of \f{supp_fig1} show the effect on the thermodynamic neuron's equilibrium activation function of including quadratic and cubic terms with the quartic coupling. The resulting activation functions display a range of forms that resemble some of those used in classical neural networks, notably the sigmoid and hyperbolic tangent functions\c{lecun2015deep,schmidhuber2015deep,dubey2022activation}.

The quartic coupling alone renders the activation function of the thermodynamic neuron nonlinear in equilibrium. However, an additional design consideration is the variance $\sigma^2$ of the neuron's equilibrium fluctuations, where 
\beq
\label{supp_var}
\sigma^2 = \av{x^2}_0-\av{x}_0^2.
\eeq
In \f{supp_fig2}(a) we plot the value of \eq{supp_var} as a function of neuron input $I$ for the quadratic case ${\bm J}=(1,0,0)$ (gray) and the quartic case ${\bm J}=(0,0,1)$ (blue). In the quadratic case the variance is constant for constant temperature, $\sigma^2=1/(2 \beta)$, reflecting the equipartition theorem. In the quartic case the variance is not constant, and is largest near $I=0$; the inset shows that its maximum value is several times that of the quadratic neuron. The mixed quadratic-quartic case ${\bm J}=(1,0,1)$, shown green in \f{supp_fig1}(c), has fluctuations at the origin comparable to the quadratic case, while the {\em mean} activation function of the case $(1,0,1)$, shown in \f{supp_fig1}(c), is similar to that of the pure quartic case. The addition of the quadratic term to the quartic one suppresses fluctuations without changing the essence of the nonlinearity. 

The larger the fluctuations of a thermodynamic neuron's output, the more samples will be required to compute a meaningful signal when observing a computer built from such neurons. For these reasons we choose our default neuron parameters to be ${\bm J}=(1,0,1)$: the quartic coupling induces nonlinearity, while the quadratic coupling serves to suppress fluctuations near $I=0$. 

The equilibrium activation function of a nonlinear thermodynamic neuron depends on temperature. In \f{supp_fig2}(b) and (c) we show the mean and variance of the equilibrium activation function of the neuron ${\bm J}= (1,0,1)$ for a range of values of $\beta$. The mean activation is nonlinear provided that temperature is not much larger than the scales of $J_2$ and $J_4$. The case in which they are comparable, i.e. $|J_2|/\kt \approx J_4/\kt \approx 1$, is acceptably nonlinear. Under the same conditions, the variance $\sigma^2$ of the equilibrium activation function, panel (b), is generally comparable in scale to the mean of the function.

\subsection{Nonequilibrium behavior} 
\label{supp_noneq}

Having assessed the equilibrium behavior of the thermodynamic neuron of \f{supp_fig1}(a), we turn to its dynamical behavior. Thermodynamic computers operate under Langevin dynamics, both overdamped and underdamped\c{aifer2024thermodynamic,melanson2025thermodynamic}. In this paper we will simulate the behavior of a thermodynamic computer using overdamped Langevin dynamics, in which case our single thermodynamic neuron evolves according to the equation\c{van1992stochastic}
\beq
\label{supp_lang1}
\dot{x}= -\mu \frac{\partial }{\partial x}U_{\bm J}(x,I) + \sqrt{2 \mu \kt} \, \eta(t),
\eeq 
where $U_{\bm J}(x,I)$ is given by~\eqq{supp_nrg}. Here $\mu$, the mobility parameter, sets the basic time constant of the computer. For the thermodynamic computers of Refs.\c{aifer2024thermodynamic,melanson2025thermodynamic}, $\mu^{-1} \sim 1$ microsecond. For damped oscillators made from mechanical elements\c{dago2021information} or Josephson junctions\c{ray2023gigahertz}, the time constant would be of order a millisecond or a nanosecond, respectively. The second term on the right-hand side of~\eqq{supp_lang1} represents the thermal fluctuations of the heat bath; $\eta$ is a Gaussian white noise satisfying $\av{\eta(t)} = 0$ and $\av{\eta(t) \eta(t')} = \delta(t-t')$. 

In \f{supp_fig3}(a)--(c) we show the dynamical evolution of the neuron $x$ under the dynamics \eq{supp_lang1}, starting from $x=0$, for various fixed values of the input $I$. Here $\beta=100$. Panel (a) shows the evolution of \eq{supp_lang1} for the case of a purely quadratic coupling, ${\bm J}= (1,0,0)$. Time traces for 11 values of $I$ are shown, evenly spaced from $-5$ to 5. The neuron is initially out of equilibrium, converging to equilibrium in about 2 time units (time is expressed in units of $\mu^{-1}$). As described by \f{supp_fig1}(b), the mean value of $x$ in equilibrium is a linear function of $I$. As described by \f{supp_fig2}(a), the size of the neuron's fluctuations in equilibrium is independent of $I$. 

\f{supp_fig3}(b) shows the case of a purely quartic coupling, ${\bm J}= (0,0,1)$. Again, time traces for 11 values of $I$ are shown, evenly spaced from $-5$ to 5; these converge at sufficiently long times to a nonlinear function $x(I)$. As described by \f{supp_fig2}(a), the fluctuations of $x$ are largest for $I=0$.

\f{supp_fig3}(c) shows the case of a mixed quadratic-quartic coupling, ${\bm J}= (1,0,1)$. The long-time mean activation function is nonlinear in $I$, and, as described by \f{supp_fig2}(a), the fluctuations near $I=0$ are suppressed relative to the purely quartic case of panel \f{supp_fig3}(b).
\begin{figure}[]
\centering
\includegraphics[width=\linewidth]{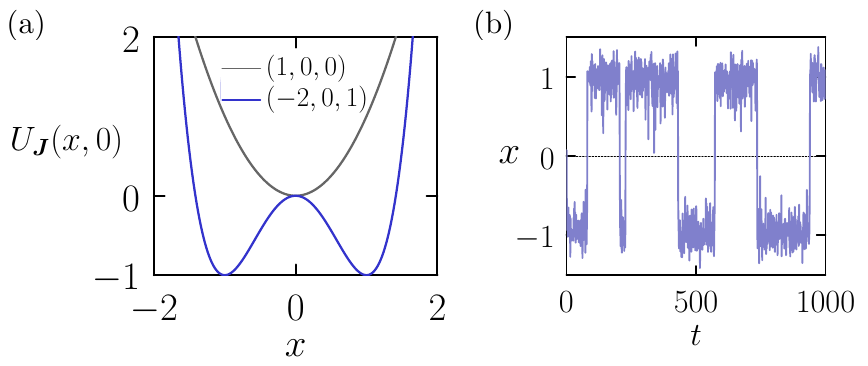}
\caption{(a) Bistable neuron potential (blue) at zero input, and (b) the resulting neuron dynamics \eq{supp_lang1} for $\beta=5$.}
\label{supp_fig_bistable}
\end{figure}

From these plots we see that for short times (here less than about 2 time units) the thermodynamic neuron's finite-time activation function does not equal the equilibrium (long-time) activation function. In \f{supp_fig3}(d) we show the value of $x$, derived from~\eqq{supp_lang1}, for a range of values of the input $I$, for various fixed values of $t$ (here we set $\beta = \infty$). For times shorter than about 0.2 units, the finite-time activation function of the neuron is linear on the scale of the inputs shown. Thus a network of such neurons would not function as a universal approximator if observed on such timescales. For longer times, however, the finite-time activation function of the neuron is nonlinear on the scale of the inputs shown. As long as we observe the output of a thermodynamic neural network on timescales longer than this threshold, it will behave as a nonlinear model. As observation time increases, the  finite-time activation function of the thermodynamic neuron converges to the long-time equilibrium activation function of the neuron, which we have designed to be nonlinear.

For certain values of the intrinsic couplings ${\bm J}$, a thermodynamic neuron can display multistability. For example, the potential~\eq{supp_nrg} with ${\bm J}=(-|J_2|,0,J_4)$ displays two minima at activations $x_0=\pm \sqrt{J_2^2/(2 J_4)}$, separated by a potential energy barrier of size $J_2^2/(4 J_4)$. One example of such a potential is shown in \f{supp_fig_bistable}(a), compared with the quadratic potential (gray). The resulting dynamics~\eq{supp_lang1} at zero input shows intermittency, as shown in \f{supp_fig_bistable}(b). Bistable neurons possess nonlinear activation functions in the vicinity of their stable states, but can also exhibit abrupt changes between states. Such bistability could be used for robust memory storage, or as a mechanism for noise-robust classification in thermodynamic neural networks. In the remainder of this paper we consider only monostable neurons.
\begin{figure}[]
\centering
\includegraphics[width=0.6\linewidth]{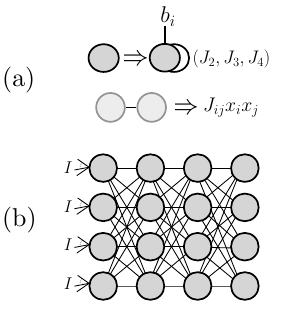}
\caption{Elements of a thermodynamic computer analogous to a neural network. (a) The thermodynamic neurons described in Figs.~\ref{supp_fig1}--\ref{supp_fig3} are connected by bilinear couplings. (b) We consider layered networks of such neurons (adjacent layers are coupled all-to-all), with total potential energy~\eq{supp_pot_tot}.}
\label{supp_fig_sketch}
\end{figure}
~
To summarize the design considerations of this section: 1) the thermodynamic neuron of \f{supp_fig1}(a) has a nonlinear equilibrium activation function if it possesses a quartic nonlinearity; 2) it is useful to also include a quadratic nonlinearity in order to suppress equilibrium fluctuations near zero neuron input; and 3) for observation times longer than some threshold, such a neuron also possesses a finite-time activation function that is nonlinear. As a result, networks of such neurons can serve as universal function approximators, both in and out of equilibrium. In the following section we illustrate these properties.

\section{Programming a thermodynamic computer}
\label{supp_net}

\subsection{A digital model of a thermodynamic universal approximator}
\label{supp_program}

Networks of nonlinear neurons are universal approximators\c{cybenko1989approximation,hornik1989multilayer}. Networks of nonlinear thermodynamic neurons are universal approximators whose operation is driven by thermal fluctuations. In this section we show that such networks can be designed on a digital computer and programmed to perform specified computations at specified observation times. 
\begin{figure*}[]
\centering
\includegraphics[width=\linewidth]{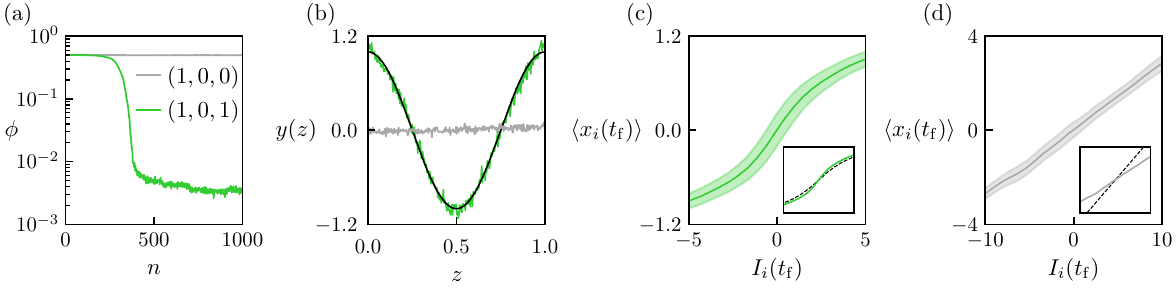}
\caption{Training a simulation model of a thermodynamic computer to express a nonlinear function at a specified observation time. (a) Loss \eq{supp_phi} as a function of evolutionary time $n$ for a layered thermodynamic computer (see \f{supp_fig_sketch}) with quadratic neurons (gray) or quadratic-quartic neurons (green). (b) Output \eq{supp_out} at observation time $\tf=1$ of the linear computer (gray) and the nonlinear computer (green), as a function of the input $z$, averaged over $M=10^3$ samples. The target function~\eq{supp_tf} is shown as a black line. (c,d). Mean neuron activations measured at observation time $\tf$ as a function of the neuron inputs at the same time, for (c) the nonlinear model and (d) the linear model. The color bands denote $\pm$ one standard deviation. Insets: the measured activation at $\tf$ compared with the {\em equilibrium} neuron activation~\eq{supp_act}.}
\label{supp_fig_sine_1}
\end{figure*}

Consider a graph of $N$ thermodynamic neurons $x_i$, with potential energy function
\beq
\label{supp_pot}
V_{\rm int}({\bm x}) = \sum_{i=1}^N U_{\bm J}(x_i,b_i)+\sum_{{\rm conn} (ij)} J_{ij} x_i x_j.
\eeq
The subscript `int' stands for `internal', and `conn' for `connections'. The first sum in \eq{supp_pot} runs over $N$ single-neuron energy terms \eq{supp_nrg}, with the intrinsic couplings of each neuron set by the vector ${\bm J} = (J_2,J_3,J_4)$. Our default choice is $(1,0,1)$. We will also consider the linear-model case $(1,0,0)$, in order to demonstrate the difference in expressive power between a linear model and a nonlinear one. The input $b_i$ to each neuron serves as a bias. 

The second sum in \eq{supp_pot} runs over all distinct pairs of connected neurons, which are determined by the graph structure imposed. We use the bilinear interaction of Refs.\c{aifer2024thermodynamic,melanson2025thermodynamic}. The computers described in those papers use an all-to-all coupling; here, to make contact with existing neural-network designs, we consider the layered structure shown in \f{supp_fig_sketch}, with all-to-all connections between layers. This design mimics that of a conventional deep fully-connected neural network. However, unlike in a conventional deep neural network, in which information flows from the input layer to the output layer, the bilinear interaction $J_{ij} x_i x_j$ ensures that neuron $i$ communicates with neuron $j$, and vice versa, and so information flows forward {\em and} backward between the layers of the thermodynamic computer.

To provide input to the thermodynamic computer we introduce the external coupling
\beq
\label{supp_pot2}
V_{\rm ext}({\bm x},{\bm I}) =\sum_{{\rm inputs} (ij)} W_{ij} I_i x_j,
\eeq
where the sum runs over all connections between the external inputs $I_i$ and the input neurons $x_j$ (here the top-layer neurons), mediated by the parameters $W_{ij}$. The total potential energy of the thermodynamic computer is then 
\beq
\label{supp_pot_tot}
V({\bm x})=V_{\rm int}({\bm x})+V_{\rm ext}({\bm x},{\bm I}).
\eeq

We assume the computer to be in contact with a thermal bath, and to evolve in time according to the overdamped Langevin dynamics
\beq
\label{supp_lang2}
\dot{x}_i= -\mu \frac{\partial V({\bm x})}{\partial x_i}  + \sqrt{2 \mu \kt} \, \eta_i(t),
\eeq 
where $V({\bm x})$ is given by \eqq{supp_pot_tot}. The Gaussian white noise terms satisfy $\av{\eta_i(t)}=0$ and $\av{\eta_i(t) \eta_j(t')} = \delta_{ij} \delta(t-t')$. 

We designate the final-layer neurons to be the output neurons. We will consider loss functions $\phi$ that are function of the outputs
\beq
\{x_i({\bm I},t)\}, i \in {\rm outputs}.
\eeq
Here ${\bm I}$ is the vector of inputs, and $x_i({\bm I},t)$ denotes the outcome of the dynamics \eq{supp_lang2} for neuron $i$ at time $t$ upon starting from zero neuron activations, ${\bm x} = {\bm 0}$ (all dynamical trajectories start from zero neuron activations). 

Because the computer is noisy, we wish to take $M$ samples of each output and average over these samples. We will consider two types of sampling. The first is reset sampling. In this mode of operation we run the computer for time $\tf$, observe the outcome, reset the neuron activations to zero, and repeat the procedure $M-1$ times, gathering $M$ samples in total. The advantage of this mode of sampling is that it naturally lends itself to parallelization: the $M$ samples can be computed independently, on distinct copies of the thermodynamic computer if such copies are available. Reset sampling can also be done using a single computer whose neurons are reset periodically. The reset-sampling average is 
\beq
\label{supp_s1}
\av{x_i({\bm I})}_{\rm r}= M^{-1}\sum_{\alpha=1}^M x^{(\alpha)}_i({\bm I},\tf),
\eeq
where the sum runs over $M$ independent realizations $\alpha$ of the dynamics \eq{supp_lang2}. The only requirement on $\tf$ is that it is long enough that the effective activation function of the neuron is nonlinear. From the considerations of \s{supp_noneq}, we set $\tf=1$ (in units of $\mu^{-1}$).

We also consider serial sampling, taking $M$ samples at intervals $t_{\rm obs}$ from a single computer running continuously for time $\tf=M t_{\rm obs}$. In this case the relevant average is
\beq
\label{supp_s2}
\av{x_i({\bm I})}_{\rm s}=M^{-1} \sum_{\alpha=1}^M x_i({\bm I},\alpha t_{\rm obs}),
\eeq
where the sum runs over $M$ samples within a single trajectory. Serial sampling is used in equilibrium in Refs.\c{aifer2024thermodynamic,melanson2025thermodynamic}. In equilibrium, a burn-in time is required for the computer to attain equilibrium, and the observation time must be long enough to obtain uncorrelated equilibrium samples. Here, by contrast, we do not require that any portion of the trajectory correspond to equilibrium (it may do, but it is not required to). As a result, we require only that $t_{\rm obs}$ be long enough for the model to be nonlinear (from the considerations of \s{supp_noneq}, we set $t_{\rm obs}=0.2$, in units of $\mu^{-1}$). No burn-in time is required, and samples do not need to be uncorrelated. 
\begin{figure*}[]
\centering
\includegraphics[width=\linewidth]{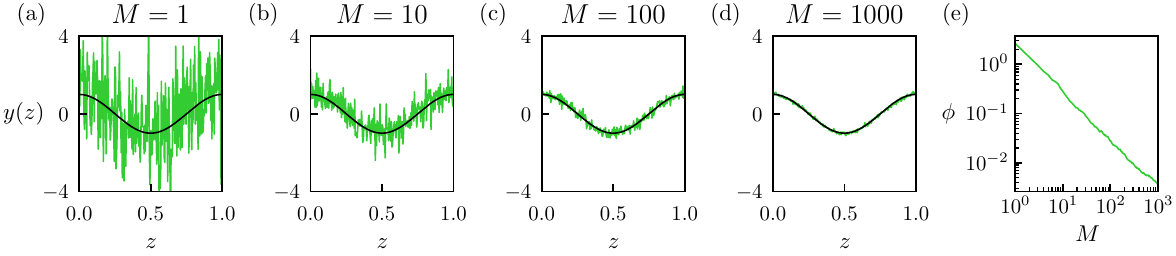}
\caption{(a--d) Output~\eq{supp_out} at time $\tf=1$ of the trained nonlinear thermodynamic computer as a function of input $z$, computed using $M$ samples. The target function~\eq{supp_tf} is shown as a black line (training was done using $M=10^3$ samples). (e) Loss~\eq{supp_phi} as a function of $M$.}
\label{supp_fig_sine_2}
\end{figure*}

\begin{figure*}[]
\centering
\includegraphics[width=\linewidth]{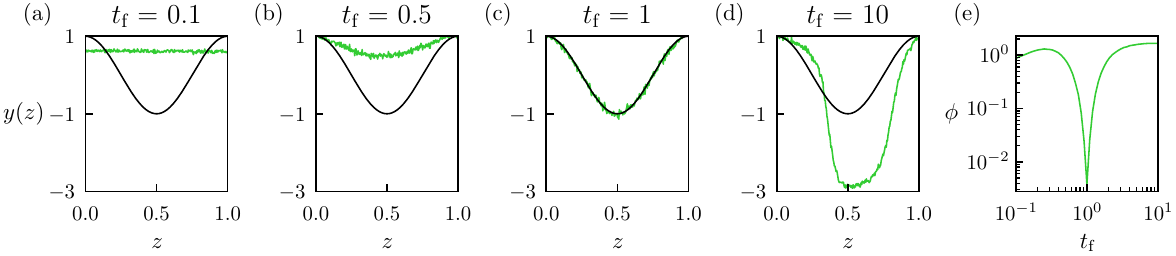}
\caption{(a--d) Output~\eq{supp_out} at various observation times $\tf$ of the trained nonlinear thermodynamic computer, as a function of input $z$, computed using $M=10^3$ samples. The target function~\eq{supp_tf} is shown as a black line. The computer is trained so that it reproduces the target function when observed at time $\tf=1$. (e) Loss \eq{supp_phi} as a function of observation time $\tf$.}
\label{supp_fig_sine_3}
\end{figure*}

\subsection{Programming the computer}

In what follows we consider the adjustable parameters of the computer to be the set ${\bm \theta}= \{W_{ij}\} \cup \{b_i\} \cup \{J_{ij}\} \cup \{f_i\}$. Here $\{W_{ij}\}$ is the set of input weights specified by \eqq{supp_pot2}; $\{b_i\}$ is the set of biases specified by~\eqq{supp_pot}; $\{J_{ij}\}$ is the set of connections specified by the same equation; and $\{f_i\}$ is a set of weights that couple to the output neurons (see \s{supp_cosine}).
\begin{figure*}[]
\centering
\includegraphics[width=\linewidth]{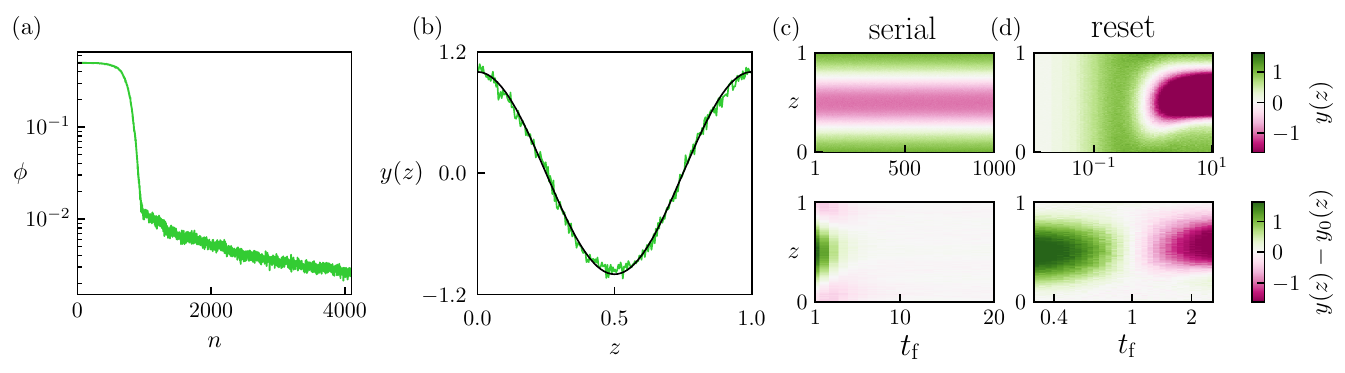}
\caption{Training a simulation model of a thermodynamic computer to express a nonlinear function in serial-sampling mode. (a) Loss \eq{supp_phi} as a function of evolutionary time $n$ for a layered thermodynamic computer (see \f{supp_fig_sketch}) with quadratic-quartic $(1,0,1)$ neurons. (b) Output \eq{supp_out} of the thermodynamic computer, evaluated using $M=10^3$ time slices, evenly spaced from $t_{\rm obs}=0.2$ to time $\tf=M t_{\rm obs}=200$ (see \eqq{supp_s2}). The black line is the target function $y_0(z)$, \eqq{supp_tf}. (c) Color maps of the output $y(z)$ of the trained computer, evaluated using $M=10^3$ evenly-spaced time slices up to time $\tf$. The bottom panel shows the difference between the computer output and the target function. The steady-state output of the computer remains close to the target function for a range of observation times. (d) Analogous plots for the computer trained in reset-sampling mode (see Figs.~\ref{supp_fig_sine_1}--\ref{supp_fig_sine_3}); here, the computer approximates the target function only at the observation time specified in training ($\tf=1$).}
\label{supp_fig_sine_4}
\end{figure*}

To program the computer we adjust the parameters ${\bm \theta}$ using a genetic algorithm instructed to minimize a loss function $\phi$. The loss is constructed from the $M$ samples \eq{supp_s1} or \eq{supp_s2} for all output neurons, and is evaluated for $K$ different sets of inputs. This evaluation requires $KM$ dynamical trajectories in reset-sampling mode, and $K$ dynamical trajectories in serial-sampling mode (loss functions are specified in \s{supp_cosine} and \s{supp_mnist}). We consider a population of $P=50$ thermodynamic computers, each of which is initialized with random parameters $\theta_i \sim {\cal N}(0,10^{-2})$. We evaluate $\phi$ for each computer, and select the 5 computers associated with the smallest values of $\phi$. These 5 are cloned and mutated to produce a new population of 50 computers. Mutations are done by adding to each parameter of each computer a Gaussian random number $\epsilon \sim C^{-1/2} {\cal N}(0,10^{-2})$. The term $C$ is the {\em fan-in}. It is equal to 1 for biases $b_i$ or weights $f_i$; it is equal to $N_j$ for weights $W_{ij}$, where $N_j$ is the number of connections entering neuron $j$; and it is equal to $(N_i + N_j)/2$ for weights $J_{ij}$. Scaling mutations by the fan-in ensures that changes to all neuron inputs are of similar scale even if the network is strongly heterogenous\c{whitelam2022training} (similar scaling ideas were used when developing efficient implementations of gradient descent\c{lecun1996effiicient}). Values of $\phi$ for the new population are calculated, the best 5 are selected, and so on.

We developed an efficient implementation of a genetic algorithm for GPU. In reset-sampling mode, each step of the genetic algorithm requires the evaluation of $PKM$ dynamical trajectories of length $\tf$ ($KM$ trajectories per computer to construct the loss function $\phi$, with $P$ computers in the genetic population). In serial-sampling mode, each step of the genetic algorithm requires the evaluation of the $PK$ dynamical trajectories of length $\tf=M t_{\rm obs}$. All trajectories can be evaluated in parallel on a set of GPUs.

Following training, the identity of the computer's parameters are fixed, and the computer can be run for any chosen input. In reset-sampling mode, testing or inference can be done with fewer than $M$ samples, if desired, as we shall describe. The parameters of the digital model of the thermodynamic computer could in principle be implemented in hardware, with the result being a device designed to output a specified computation at a specified time (or set of times, in the case of serial sampling), powered by thermal fluctuations. We note that if the hardware implementation is not an exact copy of the digital model, genetic-algorithm training could be continued directly in hardware: the procedure can be applied to an experimental system exactly as it is applied to a simulation model\c{sabattini2024adaptive}. 

\subsection{Learning a nonlinear function}
\label{supp_cosine}

To demonstrate the training and operation of a thermodynamic computer analogous to a neural network, we consider the task of expressing a nonlinear function of a single variable. In this case the computer has one input. We define the target function 
\beq
\label{supp_tf}
y_0(z) \equiv \cos( 2 \pi z),
\eeq 
and the loss function 
\beq
\label{supp_phi}
\phi  \equiv K^{-1} \sum_{j=1}^K \left(y_0(z_j)-y(z_j)\right)^2.
\eeq 
Here the sum runs over $K=250$ evenly-spaced points $z_j=j/(K-1)$ on the interval $z \in [0,1]$. The quantity
\beq
\label{supp_out}
y(z)\equiv \sum_{i \in {\rm outputs}} f_i \av{x_i(z)}_{{\rm r},{\rm s}}
\eeq
is the output of the thermodynamic computer, given the input $z$, averaged over $M$ samples (either in reset-sampling mode \eq{supp_s1} or serial-sampling mode \eq{supp_s2}). To integrate~\eqq{supp_lang2} we use a first-order Euler scheme with timestep $\Delta t = 10^{-3}$. We take $M=10^3$ for the purposes of training. The $f_i$ are parameters trained by the genetic algorithm. 

We choose a layered computer design of width 8 and depth 4. We consider two types of thermodynamic neuron: a quadratic-quartic thermodynamic neuron, ${\bm J}=(1,0,1)$, which gives rise to a nonlinear computer, and a quadratic thermodynamic neuron, ${\bm J}=(1,0,0)$, which gives rise to a linear computer. We take $\beta=10$, so that the neuron energy scale is 10 times that of the thermal energy, i.e. $J_{2,4}/\kt=10$.

In \f{supp_fig_sine_1}(a) we show the loss as a function of evolutionary time for the two models, in reset-sampling mode. The linear model fails to train -- it cannot express a nonlinear function of the input variable -- while the nonlinear model learns steadily, reaching a small value of the loss. Panel (b) shows the output functions learned by the two models: the nonlinear model has learned a good approximation of the target cosine function. The intrinsic noise of the computer is visible in the output, but for $M=10^3$ samples, for each value of $z$, the mean output signal of the computer exceeds the scale of the noise by a considerable margin.

Panels (c) and (d) of \f{supp_fig_sine_1} show the sampled neuron outputs as a function of the neuron inputs (the inputs being all signals in to the neuron, excepting the thermal noise) at the designated observation time, $\tf=1$. Consistent with the considerations of \s{supp_neuron}, the nonlinear model possesses a nonlinear finite-time activation function, indicating that we are beyond the nonlinear threshold observation time. The resulting nonlinear activation function explains the computer's ability to learn an arbitrary nonlinear function. Panel (d) confirms that the quadratic-neuron computer is at finite times a linear model, as expected.

In the inset to the two panels we show the equilibrium activation functions of the two neurons. These are qualitatively similar to their finite-time counterparts -- nonlinear or linear, respectively -- but different in detail. For the nonlinear model this difference is unimportant. Our aim is to express the target function at a finite observation time, and so what matters is that the finite-time observation time is nonlinear (which is the case if observed on timescales longer than the nonlinear threshold time). The fact that the equilibrium activation function is nonlinear simply ensures that the finite-time activation function will remain nonlinear, however long our observation time.

Training is done using $M=10^3$ samples for each value of the input $z$, but the trained computer can be used with fewer than $M$ samples if desired. In \f{supp_fig_sine_2} we show the output of the trained thermodynamic computer, as a function of the input $z$, for a range of values of $M$. The number of samples can be chosen in order to achieve a required precision.

Training in reset-sampling mode results in a thermodynamic computer programmed to express the target function at a prescribed observation time $\tf=1$. In \f{supp_fig_sine_3} we show the output of the computer at a range of observation times. The output of the computer varies as a function of time, and is equal to the target function only at the prescribed observation time. The output of the computer in equilibrium (corresponding to the long-time limit) is considerably different to the target function. In this example, therefore, the programmed thermodynamic computer operates far from equilibrium.

The model thermodynamic computer can also be trained in serial-sampling mode, as shown in \f{supp_fig_sine_4}. In this case the thermodynamic computer produces a trajectory whose stationary state approximates the target function over a range of observation times. The requirement of training is that, given an input $z$, the output of the computer, averaged over $M=10^3$ samples taken at evenly-spaced times, from $t_{\rm obs}=0.2$ to time $\tf=M t_{\rm obs}=200$, be as close as possible to the target function, $y_0(z)$,~\eqq{supp_tf}. In the top image of panel (c) we show the output of the {\em trained} computer, averaged over $M=10^3$ time slices, up to various $\tf$. The output of the trained computer remains close to the target function for a wide range of observation times. The lower image in panel (c) shows the difference between the computer output and the target function: after an initial transient (which is included in the serial-sampling average \eq{supp_s2}), the computer achieves a steady-state output that approximates the target function. This steady state may correspond to a true thermodynamic equilibrium, but we did not require this: the training requirement was only that the computer achieve a given output when averaged over a specific time interval.

The color maps in \f{supp_fig_sine_4}(d) show similar quantities for the computer trained in reset-sampling mode (see Figs.~\ref{supp_fig_sine_1}--\ref{supp_fig_sine_3}). In this mode the computer was trained to approximate the target function using $M=10^3$ samples taken at a single observation time, $\tf=1$ (see~\eqq{supp_s1}). As shown by the color plots (derived from $10^3$ samples from the trained computer taken at observation time $\tf$), the output of the computer approximates the target function only at time $\tf=1$.

\subsection{Classifying MNIST}
\label{supp_mnist}

\begin{figure*}[]
\centering
\includegraphics[width=\linewidth]{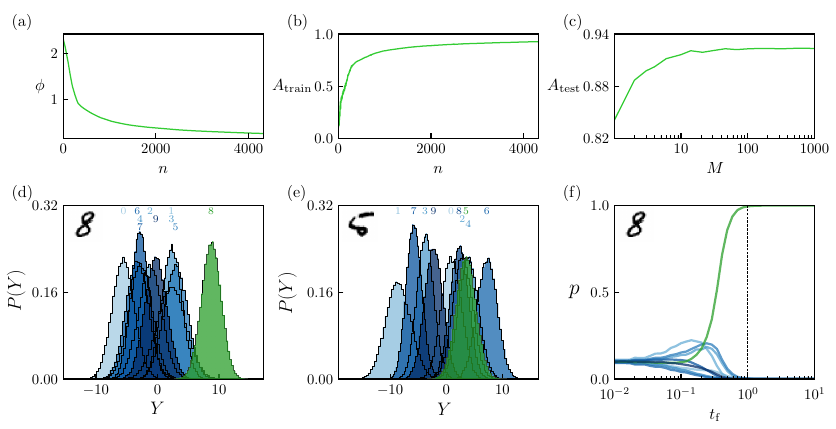}
\caption{Training a simulation model of a thermodynamic computer to classify MNIST. The computer, which consists of a 3-layer network of quadratic-quartic $(1,0,1)$ neurons, is trained in reset-sampling mode, using $M=10^3$ samples taken at observation time $\tf=1$. (a) Loss \eq{supp_mnist_loss} as a function of evolutionary time $n$. (b) Training-set classification accuracy during training. (c) Test-set classification accuracy of the trained computer, as a function of the number of samples $M$ generated by the computer (each taken at observation time $\tf=1$). (d) For a single digit, an 8, we show the probability distribution, taken over $10^5$ samples, of the computer's per-sample class predictions, \eqq{supp_labels2}. The mean value \eq{supp_labels} of each distribution, which is the value used for classification, is indicated at the top of the panel. The correct distribution is shown in green, the others in shades of blue. (e) As panel (d), but for a misclassified digit, a 5. (f) The class predictions \eq{supp_labels} of the computer, upon being shown the indicated digit, for various observation time $\tf$. The computer is trained to classify the digit at an observation time $\tf=1$ (vertical dotted line).}
\label{supp_fig_mnist}
\end{figure*}

Having confirmed the ability of a network of nonlinear thermodynamic neurons to express an arbitrary nonlinear function, we now consider a standard benchmark in machine learning, classifying the MNIST data set\c{lecun1998gradient}. MNIST consists of greyscale images of $70,000$ handwritten digits on a grid of $28 \times 28$ pixels, each digit belonging to one of ten classes $C \in [0,9]$. 

 As in \s{supp_cosine}, we take $\beta \equiv (\kt)^{-1} = 10$, so that the neuron energy scale is 10 times that of the thermal energy, i.e. $J_{2,4}/\kt=10$. We consider a 3-layer thermodynamic computer with quadratic-quartic $(1,0,1)$ neurons. Each layer has 32 neurons. Each neuron in the input layer couples to all the pixels $I_i$ of an MNIST digit via~\eqq{supp_pot2}. The output layer of 32 neurons is used to construct the computer's prediction for the class of MNIST digit ${\bm I}_j$, via the 10 \new{averaged class scores}
 \beq
 \label{supp_labels}
y^{(C)}(\bm I) \equiv \sum_{i\in {\rm outputs}}f_{i}^{(C)} \av{x_i({\bm I})}_{\rm r}.
\eeq
Here $C\in [0,9]$ is the class index, and the $f_i^{(C)}$ are 320 parameters that will be trained by genetic algorithm. Recall that the reset-sampling average is specified by~\eqq{supp_s1}.

We train the computer in reset-sampling mode, with $M=10^3$ samples taken at observation time $\tf=1$. For the loss function we choose the cross-entropy between the class probabilities predicted by the thermodynamic computer and the ground-truth labels. The probability $p^{(C)}$ that a given digit ${\bm I}_j$ is of class $C$ is obtained by applying a softmax transformation to \eq{supp_labels},
\beq
\label{supp_ml1}
 p^{(C)}({\bm I}_j) = \frac{\exp \left[y^{(C)}({\bm I}_j)\right]}{\sum_{C'=0}^9 \exp \left[y^{(C')}({\bm I}_j)\right]},
\eeq
and the cross-entropy, our loss function, is
\beq
\label{supp_mnist_loss}
    \phi = -\frac{1}{K} \sum_{j=1}^K \sum_{C=0}^9\hat{p}^{(C)}_k({\bm I}_j)\ln{p}^{(C)}_k({\bm I}_j).
\eeq
Here $\hat{p}$ is the ground-truth label, unity if ${\bm I}_j$ is of class $C$ and zero otherwise. The sum is taken over all $K=60$,000 training samples, i.e. we use full-batch learning.
\begin{figure*}[]
\centering
\includegraphics[width=\linewidth]{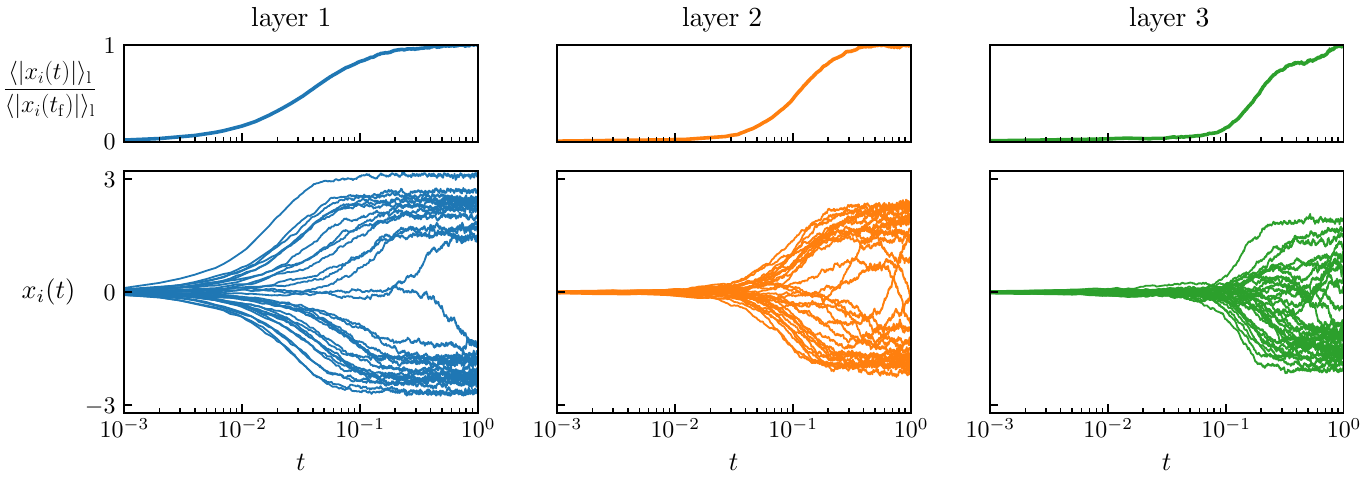}
\caption{Values of all neurons $x_i$, for a single trajectory of the 3-layer thermodynamic computer, when presented with an MNIST digit. The top panels are averaged over all neurons in the indicated layer. The computer's outputs (the layer-3 neurons) are still evolving at the observation time $\tf=1$, showing that the computer operates out of equilibrium.}
\label{supp_fig_activations}
\end{figure*}

This cross-entropy is used only in training: the expressions \eq{supp_ml1} and \eq{supp_mnist_loss} are not intended to be implemented in hardware. Once trained, the computer can be used to do classification, which we do intend to be implementable using analog hardware. Classification is done by measuring which of the 10 quantities \eq{supp_labels} is largest, when the computer is connected to a given digit. If the thermodynamic computer is realized in hardware, each of the outputs \eq{supp_labels} could be connected to a tree of comparators and a multiplexer. This additional hardware could be used to determine the identity of the quantity with the largest value, and hence the computer's predicted class for the digit.

Each evolutionary generation consists of 48 simulated thermodynamic computers, each shown 60,000 digits. Each digit is shown to each computer 1000 times, and the computer is allowed to run for time $\tf$ each time. We therefore simulate $2.9 \times 10^9$ trajectories of length $\tf$ per generation, which we parallelize over 96 GPUs. Training was done for over 4000 generations (of order 24 hours of run time), and so required the generation of more than $10^{13}$ trajectories. Training this thermodynamic computer is therefore much more expensive than training a conventional deep neural network, which can be trained to classify MNIST in seconds on a conventional computer. The advantage of the thermodynamic approach is that, once trained, the parameters  can be implemented in hardware, where the computer program will run automatically, driven only by thermal fluctuations. Conventional neural networks, once trained, must still be evaluated by explicit input of power to a CPU or GPU. Nonetheless, it will be beneficial to find less costly methods for training thermodynamic computers.

In \f{supp_fig_mnist}(a) we show the loss \eq{supp_mnist_loss} as a function of evolutionary time $n$ as the computer is trained to classify MNIST. The computer learns steadily under the action of the genetic algorithm. Panel (b) shows the corresponding training-set classification accuracy (which can be observed but is not used during training). Panel (c) shows the test-set classification accuracy of the trained computer, as a function of the number of samples $M$ (see~\eqq{supp_s1}). Training was done using $M=10^3$ samples per digit, but the trained computer can be run with considerably fewer samples ($\approx 20$) without significant loss of accuracy. Thus while the training procedure involved the generation of more than $10^{13}$ simulated trajectories, the trained computer (which is designed to be implemented in hardware) can afford to generate as few as 20 trajectories in order to classify individual digits with reasonable accuracy.

 The test-set accuracy of the trained computer is about 93\%, which is not state-of-the-art -- indeed, many other methods classify MNIST with greater accuracy\c{mnist_leaderboard} -- but the result demonstrates the ability of thermodynamic computers to do machine learning, and to carry out arbitrary nonlinear computations at specified observation times, regardless of whether or not the computer has attained equilibrium. As with conventional neural networks, better accuracy will be achieved with different computer designs and methods of training. Such incremental improvements are not the goal of the present paper. Our aim is to show proof of principle: if implemented in hardware, this thermodynamic computer would be able, powered only by thermal fluctuations, to classify MNIST digits.

In \f{supp_fig_mnist}(d) we show the output of the trained computer when presented with a single digit, an 8, which it correctly classifies. We plot the probability distribution, taken over $10^5$ samples, of the computer's per-sample predictions
 \beq
 \label{supp_labels2}
Y^{(C)}(\bm I) \equiv \sum_{i\in {\rm outputs}}f_{i}^{(C)} x_i({\bm I},\tf).
\eeq
The mean value of each distribution, corresponding to \eqq{supp_labels}, is indicated at the top of the panel. The correct distribution is shown in green, with the others in shades of blue. Panel (e) shows similar data for a digit, a 5, that is misclassified as a 6 by the computer.

\begin{figure}[]
\centering
\includegraphics[width=\linewidth]{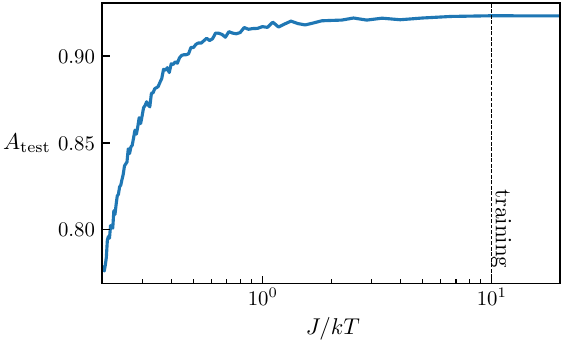}
\caption{MNIST test-set accuracy of the trained 3-layer thermodynamic computer at a range of energy scales (the parameters of the thermodynamic computer are held fixed, and the noise strength is varied). The computer is trained at the energy scale $J=10 \kt$.}
\label{supp_fig_temp}
\end{figure}

In \f{supp_fig_mnist}(f) we plot the value of the trained computer's 10 predictions \eq{supp_labels} upon being shown the indicated digit. The computer is trained to classify the digit at an observation time $\tf=1$. In this case the computer has attained a steady-state dynamics at the specified observation time, but this is not a general phenomenon: when presented with other digits, the computer's neurons are still evolving at $\tf=1$. In \f{supp_fig_activations} we show the values of all neurons $x_i$ for a single trajectory of the computer, as a function of time, when presented with an MNIST digit different to the one used in \f{supp_fig_mnist}(f). The top panels of \f{supp_fig_activations} are averaged over all neurons in the indicated layer. The computer's outputs are still evolving at $\tf=1$, showing that the computer operates out of equilibrium. 

\subsection{Role of noise in the thermodynamic computer} 

We have trained the thermodynamic computers for the cosine- and MNIST problems at an energy scale of $J_2=J_4 \equiv J=10 \kt$, which is characteristic of the energy scales of thermodynamic computing ($J \lesssim 10 \kt)$. Here, noise plays a substantial role in the dynamical evolution of the device. The energy scales of classical computing are much larger ($J \gtrsim 10^3 \kt$), allowing near-deterministic operation. 

The ability of the trained computer to operate at different noise scales depends on the nature of the problem it is trained for and its architecture. In \f{supp_fig_temp} we show the test-set accuracy of the trained thermodynamic computer at a range of energy scales (the parameters of the thermodynamic computer are held fixed, and the noise strength is varied). The computer, trained at the energy scale $J=10 \kt$, performs essentially as well when subjected to noise comparable in scale to its own energy scale ($J=\kt$). This result shows the ability of the computer to operate reliably throughout the regime characteristic of thermodynamic computing, and shows its output to be robust to small changes in noise level (which might occur if a device becomes hot during computation). For sufficiently large noise levels ($J \lesssim \kt$) the computer's performance begins to decline. 

\begin{figure}[]
\centering
\includegraphics[width=\linewidth]{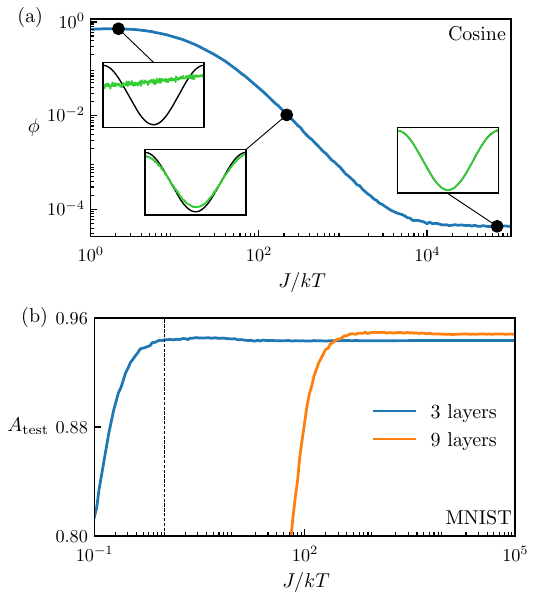}
\caption{(a) Loss associated with a simulated thermodynamic computer trained at zero noise to express a cosine function and operated at various finite noise levels. (b) MNIST Test-set accuracy of a 3-layer thermodynamic computer trained at zero noise to classify MNIST and operated at various finite noise levels (blue). The orange line shows the same thing for a 9-layer computer. The vertical black dashed line denotes the energy scale $J=\kt$.}
\label{supp_fig_zero}
\end{figure}

For larger energy scales, $J \gtrsim 10 \kt$, the computer operates reliably. This result suggests the idea of training a thermodynamic computer at larger energy scales, where noise is less important, and using the trained computer in a noisy environment. The advantage of such an approach is that, in the limit $J/\kt \to \infty$, trajectories of the computer become deterministic, and training becomes much less computationally intensive. However, upon trying this strategy we found it to work in some cases but not others.

 In \f{supp_fig_zero}(a) we show the loss associated with a thermodynamic computer trained to express the cosine function at zero noise ($J/\kt \to \infty$) and tested at a wide range of energy scales. The output values of the computer change continuously with energy scale, and so while it can tolerate the very low noise levels characteristic of classical computing, it breaks down for the larger noise levels characteristic of thermodynamic computing. In this case, the computer must be trained in the presence of noise to operate in the presence of noise.
 
In \f{supp_fig_zero}(b) we show the MNIST test-set accuracy of a 3-layer thermodynamic computer, trained at zero noise to classify MNIST, when operated at various finite noise levels (blue). This case is different to that of panel (a); here, the accuracy changes very little as the noise is increased, even up to the noise levels characteristic of thermodynamic computing. The mean outputs of the computer change continuously with noise, and so therefore does the loss, but classification accuracy depends on the relative values of the computer's outputs, not their absolute values, and so is less strongly affected. This is a convenient feature, because training is much less expensive in the deterministic zero-noise limit. There, we used backpropagation through time\c{werbos1990backpropagation} on a single trajectory of the computer (for each digit), and training takes about 30 minutes on a single GPU instead of 24 hours on 96 GPUs. For some applications, therefore, it is possible to train a thermodynamic computer at zero noise, and operate it at noise levels comparable to the energy scale of the device.

The orange line in \f{supp_fig_zero}(b) shows the same thing for a 9-layer thermodynamic computer trained at zero noise. This model is more accurate than the 3-layer one for small noise values, but fails more rapidly as noise is increased, and fails before we reach the large noise levels characteristic of thermodynamic computing. The reason for this failure is likely the diminishing signal-to-noise ratio, in the interior of the computer, as we move away from the input layer. This effect can be seen in the 3-layer computer shown in \f{supp_fig_activations}: the signal scale, and hence the signal-to-noise ratio, diminishes from input to output layers. The comparison of the the two lines in \f{supp_fig_zero}(b) indicates that robustness to noise levels different to those seen during training is an architecture-dependent phenomenon, and suggests that the layered architecture characteristic of classical neural networks may not be an optimal design for a nonlinear thermodynamic computer. 

Lastly, we note that while a thermodynamic computer trained at zero temperature may not operate reliably at finite temperature, its parameters could be used as the starting point for finite-temperature training, potentially reducing the computational expense of that process. In our finite-temperature training simulations we began with random parameters.

\end{document}